# FERMILAB PIP-II MACHINE PROTECTION SYSTEM DIGITIZED DATA NOISE ELIMINATION SCHEME AND ITS FPGA IMPLEMENTATION*

J. Wu†, A. Warner, Fermi National Accelerator Laboratory, Batavia, USA

## Abstract

In Fermilab's PIP-II machine protection system, beam loss signals from various detectors are digitized at 125 MS/s. Noise from both high-frequency sources and low-frequency 60 Hz AC power equipment can contaminate the data. To suppress noise across these ranges—especially 60 Hz and its harmonics, which overlap with beam loss signal frequencies—advanced digital processing beyond standard filtering is required. Several real-time functional blocks were simulated and tested on an FPGA: (1) a dual time-constant discharging integrator filter, (2) a de-ripple baseline extraction and storage block, and (3) a fast-recovery discharging integrator. The nonlinear IIR integrator filter removes high-frequency noise and feeds into the baseline extractor. Upon detecting abrupt beam loss, it switches to a longer time constant to prevent baseline distortion. The de-ripple block calculates a valid baseline by averaging over multiple 60 Hz periods, storing results in a 4096-word FPGA RAM. This baseline is subtracted from raw data before integration by the fast-recovery block, which resets quickly after use. All blocks achieved expected performance.

## INTRODUCTION

Fermilab's PIP-II features a brand-new, 800 MeV leading-edge Linear Accelerator (Linac) that will enable the Fermilab complex to deliver more than a megawatt of beam power to the Long Base-line Neutrino Facility (LBNF). The system includes a Warm Front-end (WFE) and a 300 meter beam transfer line to the Fermilab Booster. The WFE generates a 30 KeV H- beam, defines the beam parameters and accelerates the beam to an energy of 2.1 MeV with its RFQ and provides require bunch patterns. Capitalizing on advances in superconducting radio-frequency technology, five types of superconducting cavities will accelerate the H- ions to 800 MeV for injection into the Booster. Upgrades to the Booster, Main Injector and Recycler rings will enable them to operate at a 20 Hz repetition rate and will provide 1.2 MW proton beam for the Long Baseline Neutrino Facility (LBNF). Figure 1 shows the layout of the complex.

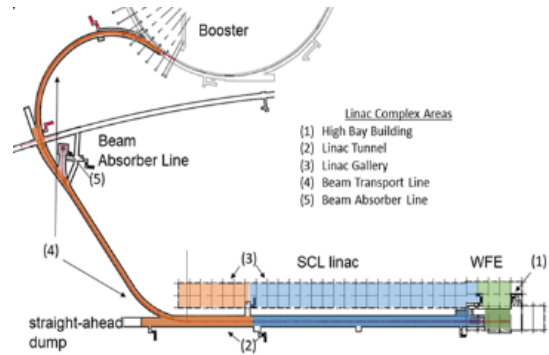

Figure 1. The layout of the PIP-II complex

The entire accelerator complex will be protected with a machine protection system (MPS)[2]. Beam current impinging on scrapers and other invasive devices and noninvasive beam current signals from ACCTs and Ring Pickups primarily in the WFE are used to determine loss thresholds and transmission losses at strategics locations in the accelerator. In the scheme for Linac protection system, the beam loss signals from these various detectors are digitized at 125 MS/s. The digitized signals are processed to extract instantaneous and differential beam loss information in the WFE along with critical operating status that ensures the accelerator complex is operating in a safe region that avoids beam and radiation damage from excessive losses. In the digitized data, noise from both high-frequency sources and low-frequency 60 Hz AC power equipment can contaminate the data, and the noise must be carefully measured and eliminated before reliable processing can be performed.

To suppress noise across these ranges—especially 60 Hz and its harmonics, which overlap with beam loss signal frequencies—advanced digital processing beyond standard filtering is required.

The block diagram of the ADC and noise elimination functional block is shown in Figure 2.

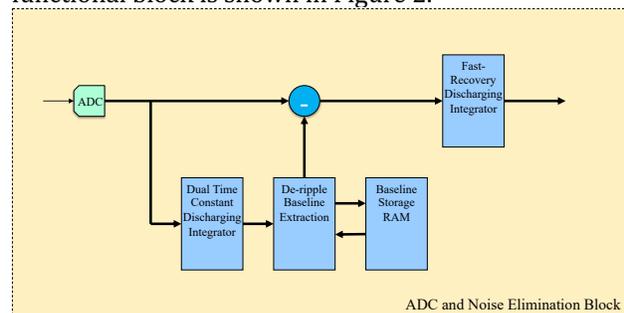

Figure 2. The diagram of the noise elimination functional blocks

The noise elimination block consists of several real-time functional blocks implemented in an FPGA serving the ADC: (1) a dual time-constant discharging integrator filter, (2) a de-ripple baseline extraction and storage block, and (3) a fast-recovery discharging integrator.

The raw data are sent through the dual time-constant discharging integrator filter and then the de-ripple baseline extraction to find the baseline. The baseline data are averaged over several 60 Hz periods and are stored in the baseline storage RAM. The baseline at the current time is subtracted from the raw data, and the high frequency noise are filtered out in the fast-recovery discharging integrator.

In this scheme, the slow variations due to 60 Hz AC noise are treated as the baseline and abrupt beam loss signals are recognized so that the baseline is maintained during the beam loss time.

In this document, details of the functional blocks are described along with the simulation results in the next section followed with conclusions.

## FUNCTIONAL BLOCKS

In the machine protection system beam loss signals from various detectors are digitized at 125 MS/s. The digitized data are sent to the processing FPGA, in which the data are fed through functional blocks one data sample per clock cycle. Therefore, the frequency of the primary system clock is 125 MHz, and it is a moderate choice suitable for mainstream FPGA families today. The functional blocks are implemented in the pipeline fashion that will help the FPGA compilers to meet the timing constraints. We will discuss the details of the functional block in the following subsections.

### The Dual Time Constant Discharging Integrator

The discharging integrator can be described with the following recursive equation:

$$y[k+1] = y[k] + A*(x[k] - y[k]) \quad (1)$$

The constant A is a ratio reflecting the response speed of the output $y[k+1]$ with the change of input $x[k]$. One can show that the step function input response of the integrator is an exponential function, and the time constant can be derived from constant A.

In typical discharging digital filters, linearity is an important requirement and therefore, its discharging time constant is a fix value which will not change with the input signal amplitude. In our application, linearity is not a required feature and therefore, the discharging time constant is allowed to change to improve the performance of the filter.

There are two contradictory situations for choosing the discharge time constant. When there is no beam loss signal, it is desirable to choose a relatively fast time constant so that the filter output tracks the variation of the baseline due to the 60 Hz AC noise. On the other hand, during the time range of the abrupt beam loss, the time constant should be relatively slow so that the output of the integrator will not deviate from the baseline too much.

The dual time-constant discharging integrator filter is a nonlinear IIR filter designed to suppress high-frequency noise and provide a clean input to the baseline extraction block. When a sudden beam loss is detected, the filter dynamically switches to a longer time constant to prevent the extraction of an incorrect baseline. The operation of the dual time-constant discharging integrator filter is shown in Figure 3.

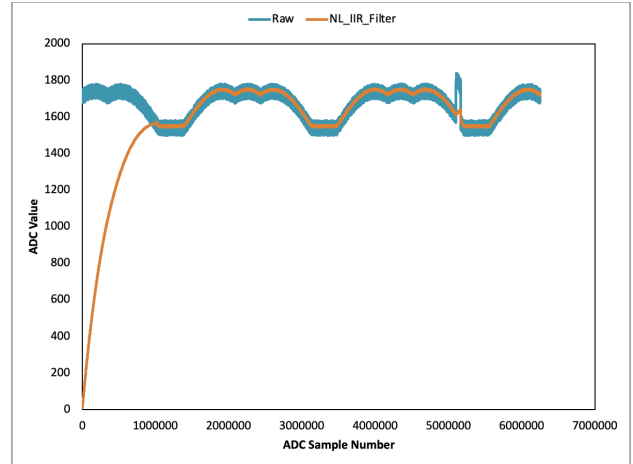

Figure 3. The operation of the dual time-constant discharging integrator filter

In the initial stage, the output of the integrator is far away from the input raw data, the internal status of the integrator is set to UNTRACKED state. In this condition, a predefined longer time constant is chosen. The output of the integrator will slowly drift to the input raw data.

Once the output is close enough to the input, the internal status is switched to TRACKED state. In this situation, a shorter time constant is used so that the output is "attracted" to the input with high frequency noise filtered out.

If a fast beam loss signal is seen, the integrator switch to the UNTRACKED state with longer time constant so that the output will not follow the input rapidly. Once the beam loss is over, the integrator will resume to TRACKED state.

The output of the integrator at the TRACKED state is sent to the next stage for de-ripple process.

### The De-ripple Baseline Extraction and Storage Block

The de-ripple block extracts a valid baseline by averaging the signal over multiple cycles of the 60 Hz period. The operation of the de-ripple block is shown in Figure 4.

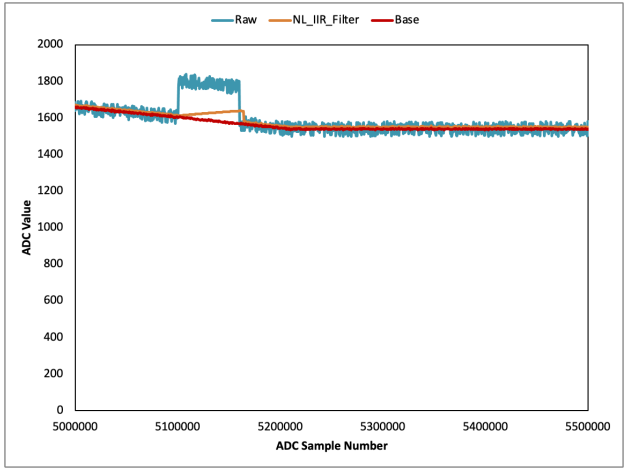

Figure 4. The operation of the baseline extraction in the de-ripple block

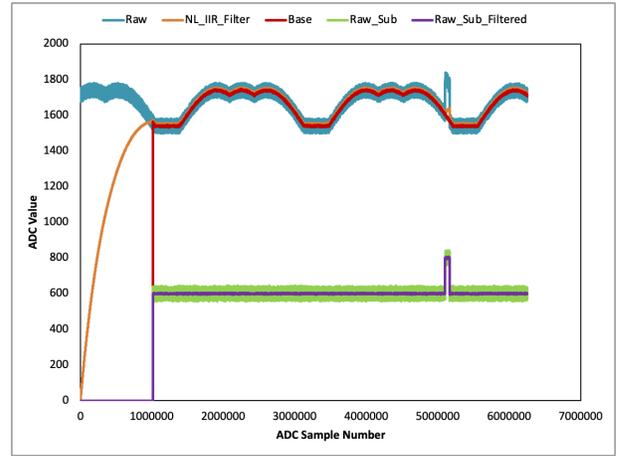

Figure 5. The operation of the fast-recovery discharging integrator

The output of the dual time-constant discharging integrator in the previous stage is the input of the de-ripple stage. A 4096-word internal RAM in the FPGA is used to store the baseline for a time interval of (1/60) second. The contents in the RAM are the average of the waveform over several periods of the (1/60Hz) and they are continuously updated as the new inputs arrive. The recursive equation of the averaged baseline is:

$$q(n+1)[m] = q(n)[m] + B*(y(n)[m] - q(n)[m]) \quad (2)$$

For each update operation, the content in the location m of RAM, $q(n)[m]$, corresponding to the current time (indexed with n) is first read out. The difference between the new input value $y(n)[m]$ and the stored old value is multiplied with a constant B and add back to $q(n)[m]$ to calculate the new value $q(n+1)[m]$. Note that the time interval of (1/60) second is represented with 4094 points. The required RAM update rate is much lower than the system clock (125 MHz), and therefore the RAM update can be designed as a multiple clock cycle process.

The constant B sets the number of previous periods of the average. For example, if B=0.5, then:

$$q(n+1)[m] = 0.5*q(n)[m] + 0.25*q(n-1)[m] + 0.125*q(n-2)[m]+\ldots \quad (3)$$

It is a weighted average with exponential weight values.

The update operation is only performed when the previous stage is in TRACKED state (and with a few other conditions to ensure the input reflects actual baseline). If the beam loss is detected, the contents of the RAM will not be updated to prevent them from being deviating away from the true baseline.

### *The Fast-recovery Discharging Integrator.*

The baseline extracted in the previous stage is then subtracted from the raw data, and the corrected signal is integrated using the fast-recovery discharging integrator. The operation of the fast-recovery discharging integrator is shown in Figure 5.

The trace "Raw_Sub" shown above represents the difference of the raw data and the baseline extracted from previous stage. (For clarity of the plot, a constant 600 is added to this value). The subtracted value is then fed into the fast-recovery discharging integrator to filter out the high frequency noise.

The operation of the integrator is essentially similar as regular integrator except the discharge time constant can be chosen for different situations. When the input is stable, a relatively longer time constant is chosen so that high frequency noise can be eliminated. But on the other hand, when a rapid beam loss is detected, the integrator chooses a shorter time constant so that the output can track the input promptly.

The actual implementation is similar as the dual time-constant discharging integrator. The switching condition between the TRACKED and UNTRACKED states and the two corresponding discharge constants are defined by the users via registers.

## CONCLUSION

Using nonlinear IIR filter structure enables separation of 60 Hz AC noise and beam loss signals which are in the same frequency range. Preliminary simulation shows that the noise elimination scheme discussed in this document functions as expected.

## REFERENCES

[1] M. Ball *et al.*, "PIP-II Conceptual Design Report," FERMILAB-DESIGN-2017-01, 2017 https://lss.fnal.gov/archive/design/fermilab-design-2017-01.pdf

[2] A. Warner *et al.*, "Experience with Machine Protection Systems at PIP2IT", in *Proc. IBIC'22*, Krakow, Poland Sep. 2022. doi:10.18429/JACoW-IBIC2022-TUP05

[3] A. Baumbaugh *et al.*, "The upgraded data acquisition system for beam loss monitoring at the Fermilab Tevatron and Main Injector," J. Instrumentation, vol. 6, 2011, DOI 10.1088/1748-0221/6/11/T11006, https://lss.fnal.gov/archive/2011/pub/fermilab-pub-11-618-ad-ppd.pdf